\documentclass[11pt,twocolumn]{article} % Specifies the document style.
\usepackage{graphicx}
\usepackage{enumerate}
\usepackage[usenames,dvipsnames]{color}
\usepackage{subfigure}
\usepackage{amsmath}
\usepackage{amssymb}
\usepackage{amsfonts}
\usepackage{color}
\usepackage[normalem]{ulem}
\usepackage{soul}
\usepackage[table]{xcolor}
\usepackage{hyperref}

\usepackage{graphicx}
\usepackage{hyperref}
\usepackage{amssymb}
\usepackage{authblk}
\usepackage{abstract}
\RequirePackage{amsmath}
\RequirePackage{fix-cm}
\RequirePackage{graphicx}
\RequirePackage{latexsym}
\RequirePackage{xspace}

\hypersetup{
    pdfnewwindow=true,      % links in new window
    colorlinks=true,        % false: boxed links; true: colored links
    linkcolor=blue,         % color of internal links
    citecolor=black,        % color of links to bibliography
    filecolor=blue,         % color of file links
    urlcolor=blue           % color of external links
}

\def\be{\begin{equation}}
\def\ee{\end{equation}}

\providecommand{\ant}{{\sc Antares}}

\providecommand{\astropart}[0]{Astropart. Phys. }
\providecommand{\apj}[0]{Astrophys. J.~}
\providecommand{\apjl}[0]{Astrophys. J. Lett. }

\providecommand{\epjc}[0]{Eur. Phys. J. C } % European Physical Journal C
\providecommand{\jcap}[0]{JCAP }
\providecommand{\ji}[0]{J. Instrum. } % Journal of Instrumentation

\providecommand{\mnras}[0]{Mon. Not. Roy. Astron. Soc. }

\providecommand{\nimA}[0]{Nucl. Instrum. Meth. A } % Nuclear Instruments and Methods in Physics Research A

\providecommand{\prd}{Phys. Rev. D. }
\providecommand{\prl}[0]{Phys. Rev. Lett. }
\providecommand{\rpp}[0]{Rep. Prog. Phys. } % Reports on Progress in Physics
\date{\today}
\author[1]{A.~Albert}
\author[2]{M.~Andr\'e}
\author[3]{M.~Anghinolfi}
\author[4]{G.~Anton}
\author[5]{M.~Ardid}
\author[6]{J.-J.~Aubert}
\author[7]{T.~Avgitas}
\author[7]{B.~Baret}
\author[8]{J.~Barrios-Mart\'{\i}}
\author[9]{S.~Basa}
% ...

\affil[1]{\scriptsize{GRPHE - Universit\'e de Haute Alsace - Institut universitaire de technologie de Colmar, 34 rue du Grillenbreit BP 50568 - 68008 Colmar, France}}
\affil[2]{\scriptsize{Technical University of Catalonia, Laboratory of Applied Bioacoustics, Rambla Exposici\'o, 08800 Vilanova i la Geltr\'u, Barcelona, Spain}}
\affil[3]{\scriptsize{INFN - Sezione di Genova, Via Dodecaneso 33, 16146 Genova, Italy}}
\affil[4]{\scriptsize{Friedrich-Alexander-Universit\"at Erlangen-N\"urnberg, Erlangen Centre for Astroparticle Physics, Erwin-Rommel-Str. 1, 91058 Erlangen, Germany}}
\affil[5]{\scriptsize{Institut d'Investigaci\'o per a la Gesti\'o Integrada de les Zones Costaneres (IGIC) - Universitat Polit\`ecnica de Val\`encia. C/  Paranimf 1, 46730 Gandia, Spain}}
\affil[6]{\scriptsize{Aix Marseille Univ, CNRS/IN2P3, CPPM, Marseille, France}}
\affil[7]{\scriptsize{APC, Univ Paris Diderot, CNRS/IN2P3, CEA/Irfu, Obs de Paris, Sorbonne Paris Cit\'e, France}}
\affil[8]{\scriptsize{IFIC - Instituto de F\'isica Corpuscular (CSIC - Universitat de Val\`encia) c/ Catedr\'atico Jos\'e Beltr\'an, 2 E-46980 Paterna, Valencia, Spain}}
\affil[9]{\scriptsize{LAM - Laboratoire d'Astrophysique de Marseille, P\^ole de l'\'Etoile Site de Ch\^ateau-Gombert, rue Fr\'ed\'eric Joliot-Curie 38,  13388 Marseille Cedex 13, France}}
% ...

\title{All-sky Search for High-Energy Neutrinos from Gravitational Wave Event
GW170104 with the \ant~Neutrino Telescope
}

\begin{document}

\onecolumn
\maketitle
\twocolumn[
%\begin{@twocolumnfalse}

%\newpage
%\begin{abstract}
\begin{abstract}
Advanced~\textsc{LIGO} detected a significant gravitational wave signal (GW170104) originating from the coalescence of two black holes during the second observation run on January 4$^{\textrm{th}}$, 2017. An all-sky high-energy neutrino follow-up search has been made using data from the \ant~neutrino telescope, including both upgoing and downgoing events in two separate analyses. No neutrino candidates were found within $\pm500$\,s around the GW event time nor any time clustering of events over an extended time window of $\pm3$\,months. The non-detection is used to constrain isotropic-equivalent high-energy neutrino emission from GW170104 to less than $\sim4\times 10^{54}$\,erg for a $E^{-2}$ spectrum.\\
\end{abstract}
%\end{abstract}
%\nopagebreak
%\end{@twocolumnfalse}
] %\twocolumn[
%\saythanks
%\cleardoublepage
%\newpage
%\newpage
%\twocolumn
\section{Introduction}\label{sec:intro}

The first two confirmed observations of gravitational waves (GWs) produced by the merger of binary black holes (BBHs) were recently made by the Advanced~\textsc{\textsc{LIGO}} interferometers during their observation run O1 \cite{gw1,gw2}. 
The second observation run of Advanced \textsc{LIGO} (O2) began in November 2016 and stopped on August 25$^\mathrm{th}$, 2017. A BBH signal, GW170104, was recorded during O2 on January 4$^{\mathrm{th}}$, 2017 at 10:11:58.6 UTC~\cite{gw3}. The false alarm rate corresponding to the signal produced by this event is less than one event over 70\,000 years. The signal was produced by the coalescence of two black holes of inferred masses of 31.2$^{+8.4}_{-6.0}$~M$_\odot$ and 19.4$^{+5.3}_{-5.9}$~M$_\odot$ at a luminosity distance of 880$^{+450}_{-390}$~Mpc. The GW source location was constrained to within 1608 deg$^2$ of the sky at 90\% credible level (region hereafter denoted as GW error box) by the LALInference reconstruction algorithm~\cite{gw3}.

Black holes with accretion disks can trigger relativistic outflows where high-energy (TeV--PeV) neutrinos (HENs) can be produced, if hadronic particles are accelerated within the jets \cite{bbhjet1,bbhjet2,bbhjet3}. Such an acceleration process can take place if magnetic fields and a long-lived debris disk remain from the stellar evolution of the black-hole progenitors or if the binary system resides in a dense gaseous environment (see e.g. \cite{Perna2016,Murase2016,Kotera2016,Bartos2017}). Since the presence of an accretion disk was not excluded in the case of GW170104, the search for muon HENs emitted before or after the merger could bring valuable information about the formation of relativistic outflows. 

The \ant\ Collaboration has joined the follow-up program of \textsc{LIGO}/Virgo detections and has received GW alerts during the whole O2 run period. The angular resolution of the \ant\ neutrino telescope ($\sim$0.4$^\circ$ at $\sim$10 TeV for muon neutrinos) compared to the size of the GW error box offers the possibility to drastically reduce the size of the region of interest in case of a coincident muon neutrino detection. 

The \ant\ field-of-view (FoV), when restricted to upgoing events, enclosed 51\% of the GW170104 error box provided by \textsc{LIGO}/Virgo at the alert time. Coincidences in time and direction between the GW signal and reconstructed muon HEN candidates were searched for in a datastream of about 1.2 events per day, selected from a total of $\mathcal{O}(100)$ upgoing neutrino track candidates triggered by ANTARES per day~\cite{TAToO_paper}. No neutrino counterpart was found and the results of this real-time analysis were transmitted via the Gamma-ray Coordinates Network (GCN) circular \#20370~\cite{GCN_circ} to the \textsc{LIGO}/Virgo follow-up community in less than 24 hours after the release of the alert. The results provided by the \ant\ Collaboration were the only real-time neutrino follow-up related to this event. The absence of neutrino candidates both temporally and spatially coincident with GW170104 allowed for deriving a preliminary upper limit on the spectral fluence emitted in neutrinos by the source at 90\% confidence level (CL). This upper limit is expressed as a function of the location of the source in equatorial coordinates and assuming a standard neutrino spectral model $dN/dE \propto E^{-2}$. This result was transmitted to the \textsc{LIGO}/Virgo follow-up community in the GCN circular \#20517~\cite{GCN_circ}.

The results of an updated high-energy neutrino follow-up of GW170104 using the \ant\ neutrino telescope are presented in this paper. The search for a transient neutrino counterpart has been extended to the full sky with different energy thresholds for events originating from below and above the \ant\ horizon, and to a larger emission timescale. The search described hereafter was performed with the most recent offline-reconstructed dataset, incorporating dedicated calibrations of positioning~\cite{antaresposition}, timing~\cite{antarestiming} and efficiency~\cite{ANTARES2}. The analysis has been optimized to increase the sensitivity of the detector at the time of the alert. Two neutrino spectral models were assumed: a generic model $dN/dE=\phi_0 E^{-2}$ typically expected for Fermi acceleration and a model with a high-energy cutoff $dN/dE=\phi_0 E^{-2}\mathrm{exp}\left[-\sqrt{(E/100\mathrm{TeV})}\right]$. The second model is expected for sources with exponential cutoff in the primary proton spectrum~\cite{Kappes2007}. Finally, systematic errors affecting the corresponding upper limits on neutrino emission are accounted for. 

The capabilities of the \ant\ detector and the search procedures are summarized in Section \ref{sec:search}. The constraints on the neutrino fluence and total energy emitted in neutrinos derived from the non-detection of a neutrino counterpart for GW170104 are presented in Section \ref{sec:results}. The conclusions are reported in Section \ref{sec:conclusion}.

\section{High energy neutrino search}
\label{sec:search}

\ant~\cite{ANTARES} is an underwater neutrino telescope located in the Mediterranean Sea, offshore Toulon (France). It is composed of an array of photomultiplier tubes (PMTs), anchored at a depth of 2475 m under the sea level. Neutrinos with energies above $\sim10^2$~GeV are detected through the Cherenkov light induced by relativistic particles created from the interaction of neutrinos with matter. In addition to astrophysical neutrino signals, both atmospheric muons and neutrinos can lead to detectable light in the detector and are considered as background events. However, only neutrinos can traverse the Earth. Looking at upgoing particles in the detector reference frame allows for removing a large part of the downgoing atmospheric muon background. Remaining mis-reconstructed downgoing muons are further rejected by applying cuts on the reconstruction quality parameters. In addition, the intense background of downward going atmospheric muons is drastically reduced by the requirement for a joint time and space coincidence with the GW time and spatial error box. This allows for searching for a neutrino counterpart for GW170104 in both upgoing and downgoing datasets, which consist of events originating respectively from below and above the \ant\ horizon.

\begin{figure}[h!]
\begin{center}
\resizebox{0.49\textwidth}{!}{\includegraphics[trim = 1cm 3cm 1cm 3cm, clip]{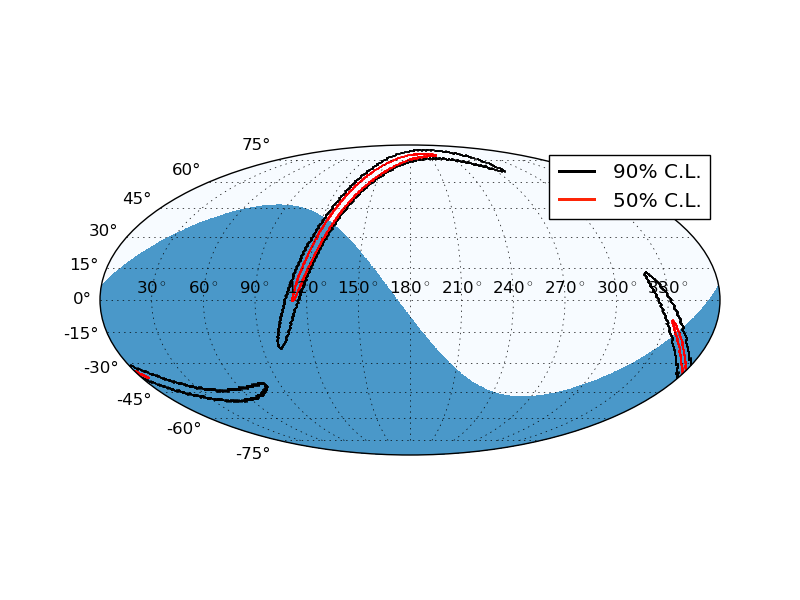}}
\end{center}
\caption{Visibility map of GW170104 in equatorial coordinates. The sky regions below and above the \ant\ horizon at the alert time are shown in blue and white respectively. Events that originate from the blue (white) region will be seen as upgoing (downgoing) in the detector frame. The red and black contours show the reconstructed probability density contours of the GW event at 50\% and 90\% credible level respectively.}
\label{fig:skymap}
\end{figure}

Considering the refined location probability provided by the \textsc{LIGO}/Virgo LALInference software~\cite{LALInf_paper}, there is a 52\% chance that the GW emitter was below the \ant\ horizon where any neutrino events from this part of the sky would be seen as upgoing in the detector frame. This corresponds to a 45\% probability for the source to be located inside the GW error box and below the \ant\ horizon (see Fig. \ref{fig:skymap}). To extend the overlap between the \ant\ FoV and the GW error box, downgoing events have been added to the search in an independent analysis.

All-sky \ant\ data have been searched for track events produced by $\nu_\mu$ and $\bar{\nu}_\mu$ charged current interactions coincident with GW170104 using a time window of $\pm 500$\,s around the GW transient (Sections \ref{Southern_search} and \ref{Northern_search}). This time window was adopted as the standard search window for previous joint GW-HEN searches~\cite{Baret2011}, for instance in the case of GW150914 and GW151226~\cite{GW150914_paper,GW151226_paper}. A search for a neutrino counterpart within an extended time window of $\pm 3$ months has also been done using the online datastream of \ant\ (Section \ref{extended_search}).

\subsection{Search below the \ant\ horizon}\label{Southern_search}
A binned search for coincident upgoing neutrinos was performed following a blind procedure. The track reconstruction algorithm computes both the neutrino direction, together with an estimated error $\beta$, and a quality parameter $\Lambda$~\cite{PSpapers_acc}. This sample is dominated by background events from mis-reconstructed downgoing atmospheric muons, which deposit energy in the detector through stochastic processes. The dataset was reduced by adjusting $\Lambda$ such that any event passing the search criteria and located within the GW error box, below the \ant\ horizon, would lead to a detection with a significance level of $3\sigma$. This optimization was carried out on data outside the 1000\,s time window used in this search. A Monte Carlo simulation of the detector response~\cite{2013NIMPA.725...98M,2016EPJWC.11602002F} at the alert time allows for estimating the relative contribution of the atmospheric neutrinos and
the mis-reconstructed atmospheric muons to the background rate below the \ant\ horizon and within $\pm500$\,s. A total of $2.2\times10^{-2}$ atmospheric neutrino candidates are expected while the number of mis-reconstructed downgoing muons amounts to $3.7\times10^{-2}$ events over 2$\pi$~sr. 

After unblinding of the dataset, no event temporally coincident with GW170104 was found.

\subsection{Search above the \ant\ horizon}\label{Northern_search}
A search for coincident neutrino candidates detected above the \ant\ horizon was carried out by selecting downgoing events with $\beta$ smaller than 1$^\circ$. The cuts are optimized on a combination of $\Lambda$ and the number of hits used in the reconstruction, where a hit corresponds to a PMT signal above a given threshold. The number of hits can be considered as a proxy of the muon/neutrino energy. Indeed, downgoing atmospheric muons are less likely to produce a number of hits as large as that produced by very high-energy cosmic neutrinos. Anew, the selection criteria were optimized such that one event occurring within the signal time window of 1000\,s and located inside the GW error box located above the \ant\ horizon would lead to a detection with a significance level of $3\sigma$. The final set of cuts is chosen as the one maximizing the fraction of surviving signal events. The final sample is mostly composed of atmospheric muons with a total of $8.2\times10^{-2}$ background events expected above the \ant\ horizon within $\pm500$\,s. The median neutrino energy that would be detected by \ant\ for a $E^{-2}$ signal spectrum is about a factor of 10 higher for this analysis compared to the search below the horizon described in Section \ref{Southern_search}. 

After unblinding of the dataset, no event temporally coincident with GW170104 was found.

\subsection{Extended time window search}\label{extended_search}

The time window of $\pm 500$\,s was chosen by assuming that if compact binary mergers are related to gamma-ray bursts then the neutrino signal should occur close in time to the GW emission. This time window is large enough to catch potential precursor neutrino emission and time offsets with respect to the GW signal~\cite{Baret2011}. For completeness, to probe non-standard propagation scenarios similar to those described in \cite{GRB_stack}, a search for shifted and/or longer-lasting emission over $\pm 3$~months around the GW alert was performed by looking for time clustering of upgoing neutrino events.

The events selected from the online datastream used in the \ant\ real-time alert program~\cite{TAToO_paper} are investigated for time clustering. The spatial clustering of the events and their coincidence with the GW error box were investigated \textit{a posteriori}.

An unbinned likelihood search was performed following the methodology applied in previous analyses~\cite{time_dept1,time_dept2}. For each combination of two events $a$ and $b$, a signal probability for the $i^{\mathrm{th}}$ event is defined as:
\begin{equation}
S^{a,b}_i=\frac{H(t_{b}-t_i)H(t_i-t_{a})}{t_{b}-t_{a}},
\end{equation}

\noindent with $H$ the Heaviside function\footnote{$H(0)$ is defined as 1.} and $t_{a}$ and $t_{b}$, the detection time of events $a$ and $b$ (with $t_{a}$ $<$ $t_{b}$). The background time probability for the $i^{\mathrm{th}}$ event, $B_i$, is derived directly from the probability density function (PDF) of the downgoing reconstructed events. In this way, the background distribution reflects the evolution of the event rate due to the variability of the data taking conditions. 

Given a dataset of $N$ events, the likelihood function $\mathcal{L}^{a,b}(n_s)$ for a given pair of events occurring at times $t_{a}$ and $t_{b}$ is defined as:

\begin{equation}
\mathcal{L}^{a,b}(n_s) = \prod_{i=1}^{N}\left[\frac{n_s}{N}S^{a,b}_i + \left(1-\frac{n_s}{N}\right)B_i\right],
\end{equation}

\noindent where $n_s$ is the unknown number of signal events. For each pair of events occurring at times $t_{a}$ and $t_{b}$, the likelihood is maximized with respect to $n_s$ to provide the best-fit number of events $\hat{n}_s$. The test statistic (TS$^{a,b}$) is computed from the likelihood ratio of the background-only (null) hypothesis over the signal-plus-background hypothesis as:

\begin{equation}
\mathrm{TS}^{a,b} = -2\cdot\mathrm{log}\left[\frac{T}{t_{b}-t_{a}}\frac{\mathcal{L}(n_s=0)}{\mathcal{L}^{a,b}(\hat{n}_s)}\right],
\end{equation}

\noindent where the term $\frac{T}{t_{b}-t_{a}}$ in the square brackets is a trial factor. This quantity is needed to correct for the fact that there are many more independent small time windows than large ones, which tend to favour very short flares. The parameter $T$ corresponds to the dataset livetime.
Given a sample of $N$ events, $\frac{N(N-1)}{2}$ values of TS were computed (one for each pair of events $a$ and $b$). The cluster that maximizes the TS is finally considered as the most significant one.\\

To compute the $p$-value of the most significant cluster, $\mathcal{O}$(10\,000) pseudo-experiments were generated, each of them consisting of $N$ events drawn randomly from the time PDF of the background. The fraction of trials for which the TS value is larger than the one obtained from the data is referred to as the $p$-value.

The most significant time cluster has been found to contain $\hat{n}_s$=8.3 fitted signal events, occurring between $t_{min} (MJD)$=57682.73398 and $t_{max} (MJD)$=57685.62900 ($t_{max}-t_{min}=$2.89 days). It leads to a post-trial $p$-value of 70\% and is thus consistent with the background-only hypothesis. In addition, the \ant\ events contained in this time window are not spatially compatible and do not overlap with the GW error box.

\section{Astrophysical constraints}
\label{sec:results}

The non-detection of joint GW and neutrino signals is used to constrain neutrino emission from the GW source. Upper limits on both the fluence and the total energy emitted in neutrinos are presented in the form of skymaps since the sensitivity of \ant\ depends on the source direction. 

\subsection{Constraints on the neutrino spectral fluence}\label{sec:constraints_astro}

Upper limits at 90\% CL on the neutrino fluence from a point source within $\Delta t=\pm500$\,s were calculated using the null result and the detector acceptance, estimated via a Monte Carlo simulation of the detector response at the time of the GW signal. This simulation is produced on a run-by-run basis~\cite{2013NIMPA.725...98M,2016EPJWC.11602002F} to account for the variation of the data taking conditions under the sea. The two spectral models described in Section \ref{sec:intro} were considered.

The number of neutrino events expected to be observed by \ant\ from a point source at declination $\delta$ and with a neutrino flux $\mathrm{d}N/\mathrm{d}E$ (in GeV$^{-1}$~cm$^{-2}$~s$^{-1}$) in a time window $\Delta t$ is given by:

\begin{equation}
N_\mathrm{events}=\Delta t\int\frac{\mathrm{d}N}{\mathrm{d}E}(E)\mathrm{A}_\mathrm{eff}(E,\delta)\mathrm{d}E,
\end{equation}

\noindent where A$_\mathrm{eff}(E, \delta)$ is the effective area of \ant\ at the alert time which take into account the absorption of neutrinos by the Earth and depends on the neutrino energy $E$, the source declination $\delta$ and the applied cuts.

The upper limit on the fluence, $\phi_0^{90\%}$, is defined as the fluence value that on average would produce 2.3 detected neutrino events. Assuming a $dN/dE=\phi_0E^{-2}$ neutrino spectral model, it is derived as:

\begin{equation}
\phi_0^{90\%} = \frac{2.3}{\Delta t \int E^{-2}\mathrm{A}_\mathrm{eff}(E_\nu,\delta)\mathrm{d}E},
\label{eq:phi0}
\end{equation}

\noindent where the denominator refers to the instantaneous acceptance of the \ant\ detector at the time of the alert computed between 1~GeV and 100~PeV. Equation \ref{eq:phi0} also applies for the second spectral model considered in this study. The same methodology is used to derive $\phi_0^{90\%}$ in the case of the second considered spectral model and on the search above the horizon.

\begin{figure}[h!]
\begin{center}
\resizebox{0.49\textwidth}{!}{\includegraphics{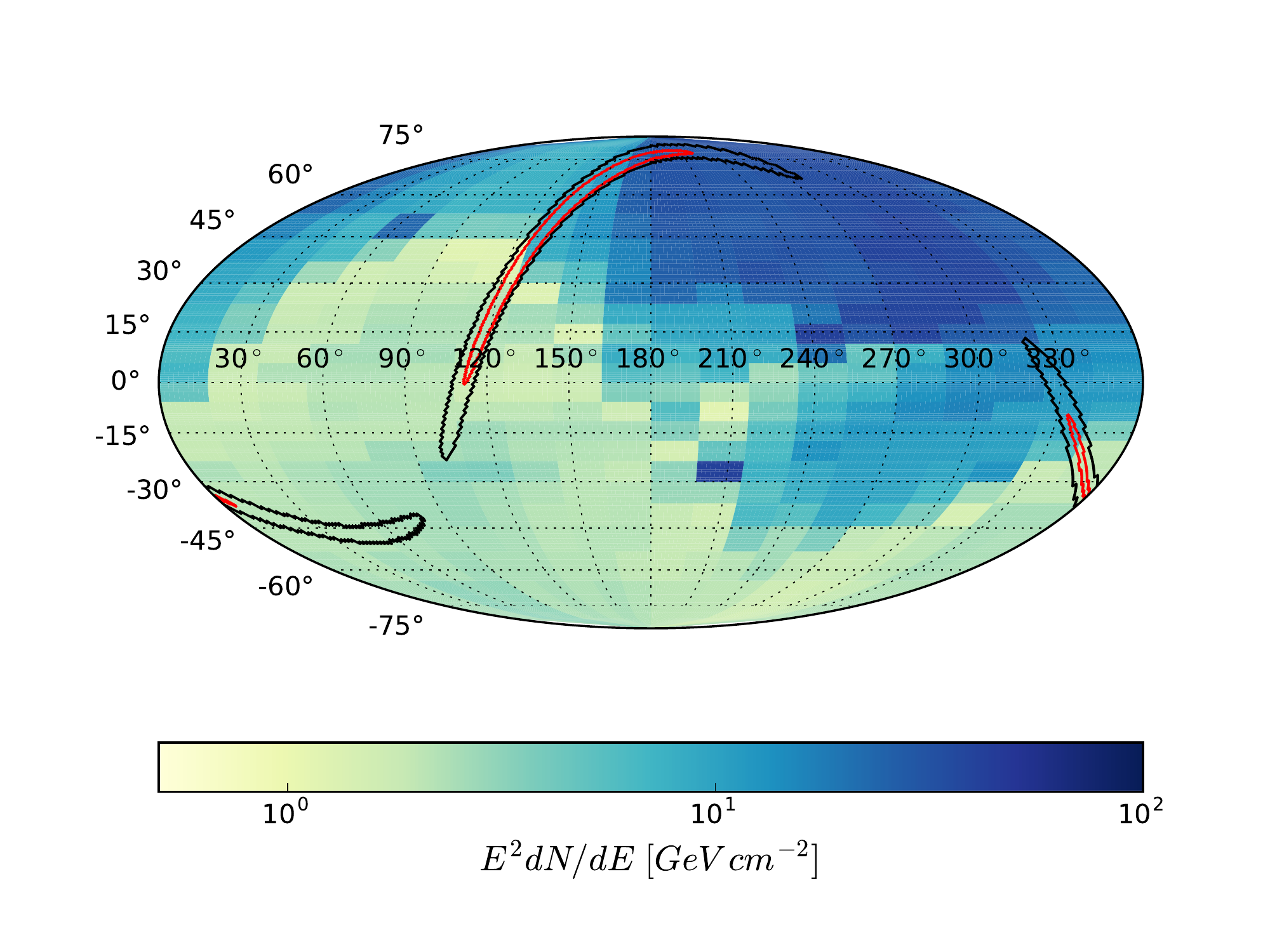}}
\resizebox{0.49\textwidth}{!}{\includegraphics{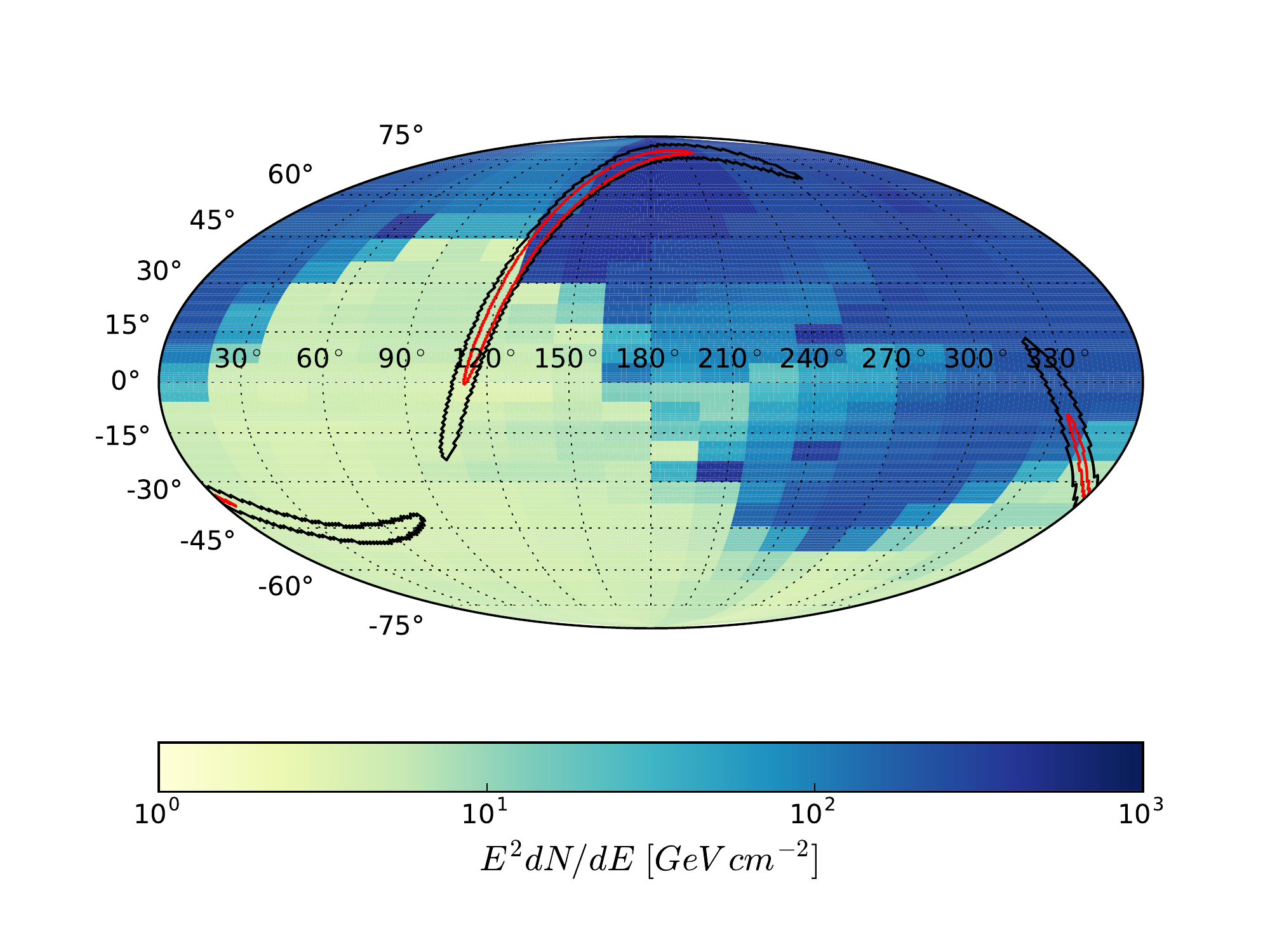}}
\end{center}
\caption{All-sky upper limit on the neutrino spectral fluence ($\nu_\mu + \bar{\nu}_\mu$) from GW170104 as a function of source direction assuming $\mathrm{d}N/\mathrm{d}E\propto E^{-2}$  (top) and $\mathrm{d}N/\mathrm{d}E\propto E^{-2}\mathrm{exp}\left[-\sqrt{(E/100\mathrm{TeV})}\right]$ (bottom) neutrino spectra. The red and black lines show the GW skymap contours at 50\% and 90\% credible levels, respectively. Skymaps are defined in equatorial coordinates.}
\label{fig:UL1}
\end{figure}

Fig. \ref{fig:UL1} shows the neutrino spectral fluence upper limit ($\phi_0^{90\%}$) for GW170104 as a function of the source direction for both spectral models. Computed from the Monte Carlo simulation, the energy range corresponding to the 5\%--95\% quantiles of the neutrino flux below the \ant\ horizon is equal to [3.2~TeV; 3.6~PeV] for the $dN/dE=\phi_0 E^{-2}$ spectral model and [1.4~TeV; 270~TeV] for the model with exponential cutoff at 100 TeV. Above the \ant\ horizon, the 5\%--95\% quantiles of the neutrino flux are respectively equal to [120~TeV; 22~PeV] and [53~TeV; 950~TeV].

The systematic uncertainties on the fluence upper limits have been estimated by summing quadratically \textit{i)} the systematic error on the acceptance of the detector and \textit{ii)} the uncertainty related to the ability of the \ant\ run-by-run Monte Carlo approach to accurately reproduce the variable data taking conditions on short time scales. This latter effect can become dominant when looking for transient neutrino sources. 

The systematic error on the acceptance has been computed by reducing the efficiency of each optical module by 15\% in the detector simulations. This leads to a 15\% uncertainty on the acceptance as detailed in \cite{PSpapers_acc}.

The second source of uncertainty has been constrained by quantifying the ability of the run-by-run Monte Carlo to accurately reproduce the evolution of the event rate observed in the data from one run to another. Due to the low number of events passing the optimized quality cuts, the run-by-run agreement between data and Monte Carlo can only be assessed by loosening the cuts, on a data sample dominated by atmospheric muon events. The variations of the muon rate are well reproduced with a median relative variation between data and Monte Carlo smaller than 20\%. In addition, the short-timescale fluctuations of the event rate are smaller for signal neutrinos with a $E^{-2}$ spectral model than for atmospheric muons. This was expected since the detector geometry is optimized for upgoing events. Thus, the systematic error of 20\% is considered a conservative value.

The resulting total systematic uncertainty on the fluence upper limits, which applies for both upgoing and downgoing event searches, is 25\%.

\subsection{Constraints on the total energy emitted in neutrinos}

The GW signal contains also information on the source distance which can be reconstructed around the GW error box~\cite{Singer2016}. This information can be used to derive an upper limit on the total energy radiated in neutrinos as a function of the direction as performed in \cite{GW151226_paper}. 

The most likely value $D(\vec{x})$ of the distance for each direction $\vec{x}$ is used to calculate the upper limit on the total isotropic-equivalent energy emitted in neutrinos by the source as:

\begin{equation}
E^{\mathrm{UL}}_{\nu, \mathrm{iso}}\left(\vec{x}\right) = 4\pi\left[D(\vec{x})\right]^2\int\frac{\mathrm{d}N}{\mathrm{d}E}\left(E, \vec{x}\right)E\mathrm{d}E.
\end{equation}

\begin{figure}[h!]
\begin{center}
\resizebox{0.48\textwidth}{!}{\includegraphics{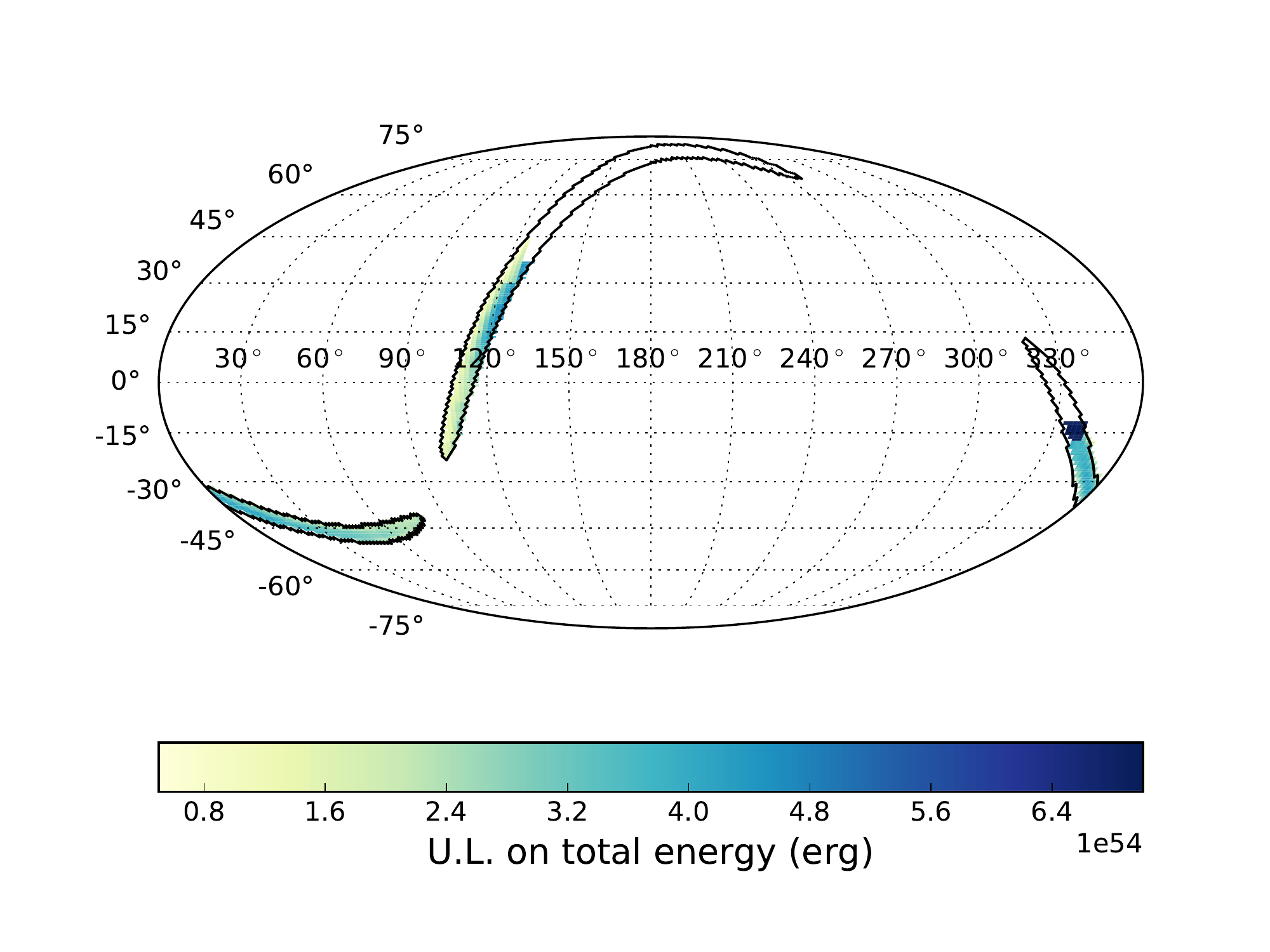}}
\resizebox{0.48\textwidth}{!}{\includegraphics{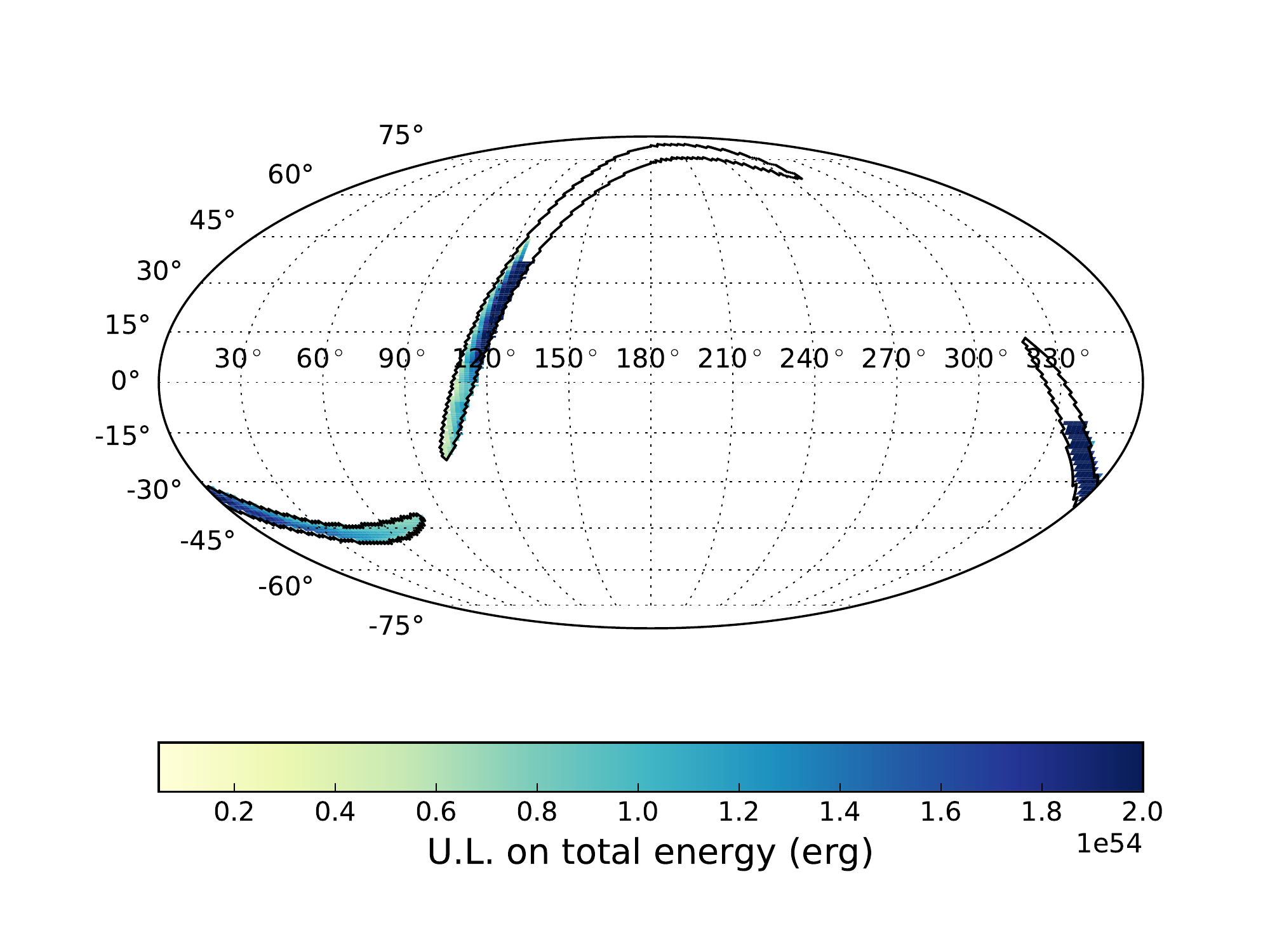}}
\end{center}
\caption{Upper limits on the total energy radiated in $\nu_\mu + \bar{\nu}_\mu$ from GW170104 as a function of source direction assuming $\mathrm{d}N/\mathrm{d}E\propto E^{-2}$ (top) and $\mathrm{d}N/\mathrm{d}E\propto E^{-2}\mathrm{exp}\left[-\sqrt{(E/100\mathrm{TeV})}\right]$ (bottom) neutrino spectra. The upper limits are given for the sky below the \ant\ horizon, where they are the most stringent. Skymaps are defined in equatorial coordinates.}
\label{fig:energyUL}
\end{figure}

Upper limits on the total energy are computed for both $\mathrm{d}N/\mathrm{d}E\propto E^{-2}$ and $\mathrm{d}N/\mathrm{d}E\propto E^{-2}\mathrm{exp}\left[-\sqrt{(E/100\mathrm{TeV})}\right]$ neutrino spectral models. The spectrum is integrated over the range $\left[100 \mathrm{~GeV}; 100 \mathrm{~PeV}\right]$. The upper limits as a function of source direction are shown in Fig. \ref{fig:energyUL} for the region corresponding to upgoing events for \ant.  It can be seen in Fig. \ref{fig:energyUL} that the derived constraints depend on the position on the sky as both the fluence upper limits and the distance constraints do. The 5\%--95\% range of values is [$1\times10^{54}$; $4\times10^{54}$]~erg and [$6\times10^{53}$; $4\times~10^{54}$]~erg, for the $E^{-2}$  and the 100 TeV cutoff models respectively. The strongest constraint is obtained at declination $\delta\sim-17^\circ$ with $E<5\times10^{53}$~erg for a $E^{-2}$ spectrum and $E<3\times10^{53}$~erg for the spectral energy distribution with cutoff at 100~TeV.  These values are about 10\% of the total energy of $\sim3.6\times 10^{54}$~erg emitted from GW170104 in gravitational waves. The results obtained for downgoing events are not provided since the neutrino fluence upper limits are much weaker. The uncertainty on the total energy emitted in neutrinos is estimated by accounting for both the systematic error on $\phi_{0}^{90\%}$ (computed above) and the 1$\sigma$ standard deviation on the distance provided by the \textsc{LIGO}/Virgo GW event reconstruction, leading to an average value of $\sim40$\%.
\section{Conclusion}\label{sec:conclusion}
No neutrino emission associated with the third confirmed binary black hole merger, GW170104 was detected in the \ant\ data. This non-detection was used to derive an upper limit to the total neutrino emission from GW170104 of $\sim 4\times$10$^{54}$ erg, for a generic $E^{-2}$ neutrino spectrum and for a high-energy cutoff spectrum $E^{-2}\mathrm{exp}\left[-\sqrt{(E/100\mathrm{TeV})}\right]$ as expected for sources with an exponential cutoff in the proton spectrum. These results are of the same order of magnitude as the ones previously published for GW150914, LVT151012 and GW151226. The strongest constraint, obtained at a declination $\delta\sim-17^\circ$, shows that if the GW source is located at this position on the sky, the total energy emitted in neutrinos from GW170104 is not more than 10\% of the total energy emitted in gravitational waves.\\ \\

The \ant\ Collaboration is grateful to the LIGO Scientific Collaboration and the Virgo Collaboration for the setting up of an impressive follow-up observation program, and for sharing invaluable scientific information for the benefit of the emerging multi-messenger astronomy.
The authors acknowledge the financial support of the funding agencies:
% France:
Centre National de la Recherche Scientifique (CNRS), Commissariat \`a
l'\'ener\-gie atomique et aux \'energies alternatives (CEA),
Commission Europ\'eenne (FEDER fund and Marie Curie Program),
Institut Universitaire de France (IUF), IdEx program and UnivEarthS
Labex program at Sorbonne Paris Cit\'e (ANR-10-LABX-0023 and
ANR-11-IDEX-0005-02), Labex OCEVU (ANR-11-LABX-0060) and the
A*MIDEX project (ANR-11-IDEX-0001-02),
R\'egion \^Ile-de-France (DIM-ACAV), R\'egion
Alsace (contrat CPER), R\'egion Provence-Alpes-C\^ote d'Azur,
D\'e\-par\-tement du Var and Ville de La
Seyne-sur-Mer, France;
% Germany: 
Bundesministerium f\"ur Bildung und Forschung
(BMBF), Germany; 
% Italy
Istituto Nazionale di Fisica Nucleare (INFN), Italy;
% Netherlands
Nederlandse organisatie voor Wetenschappelijk Onderzoek (NWO), the Netherlands;
% Russia
Council of the President of the Russian Federation for young
scientists and leading scientific schools supporting grants, Russia;
% Romania
National Authority for Scientific Research (ANCS), Romania;
% Spain 
Mi\-nis\-te\-rio de Econom\'{\i}a y Competitividad (MINECO):
Plan Estatal de Investigaci\'{o}n (refs. FPA2015-65150-C3-1-P, -2-P and -3-P, (MINECO/FEDER)), Severo Ochoa Centre of Excellence and MultiDark Consolider (MINECO), and Prometeo and Grisol\'{i}a programs (Generalitat
Valenciana), Spain; 
% Marocco
Ministry of Higher Education, Scientific Research and Professional Training, Morocco.
% A.O.B.:
We also acknowledge the technical support of Ifremer, AIM and Foselev Marine
for the sea operation and the CC-IN2P3 for the computing facilities.
%#################################################################

%]  %\twocolumn[
\end{document}